\documentclass[11pt]{article}
\usepackage{epsfig}  
\begin{document}
\noindent
\begin{center}{ \large \bf About new quasilinear model of heat conduction with finite velocity
of heat front movement. }
\end{center}

\begin{center}{ \large \bf A.N. Skripka }
\end{center}

\begin{center}{  KCK Soft, Kiev, Ukraine }
\end{center}

\begin{center}{ \small skripka@ukrpost.net }
\end{center}

{ \it A new quasilinear mathematical model of heat conduction with finite velocity
of heat front movement is offered.
}

\vspace{3mm}

{ \bf 1. The analysis of existing models of heat conduction. }

Now as basic mathematical model in the theory of heat conduction the linear
parabolic differential equation is used. In case of one spatial dimensional:
$$
\frac{\partial T}{\partial t}=D_T\frac{\partial^2T}{\partial^2x}.\eqno(1)
$$
Also it can be presented as system of two equations:
$$
c\rho\frac{\partial T}{\partial t}=-div\vec{J},\eqno(2)
$$
$$
\vec{J}=-\vec{i}\lambda\frac{\partial T}{\partial x}.\eqno(3)
$$
The following designations are entered: $ T(t,x) $ - field of temperature,
$ \vec{J}(t,x) $ - vector of heat flow, $ \lambda $ - coefficient of heat conduction,
$ D_T=\lambda/(c\rho) $ - coefficient of temperature conduction, $ \rho $ - density, 
$ c $ -  thermal capacity, $ div = \partial /\partial{x} $, 
$ \vec{i} $ - unit vector.

Equation (1) is obtained taking into account suppositions about 
instantaneous relaxation of the heat flow, that contradicts physical 
conceptions about the nature of heat. This leads to infinite velocity 
of front heat movement in the given model.

Sometimes model the nonlinear heat conduction equation:
$$ \frac{\partial T}{\partial t}=\frac{\partial}{\partial x}
  (D_T(T)\frac{\partial T}{\partial x}),\eqno(4)
$$
where the coefficient of heat conduction is function of temperature,
or linear hyperbolic equation Cattaneo is used. But both these models
have basic lacks[2], which do not allow to widely use them at the decision of practical tasks.

In this work new heat conduction equation is obtained, which is a generalization 
of the linear equation (1).
For this purpose equation (1) is transformed to the nonlinear equation in such a manner, 
that quantitatively its solutions differ from solutions of equation (1) a little,
but the new equation will describe the process of heat conductivity qualitatively precisely.

In the new equation there is a parameter, connected with velocity of heat conduction front.

\newpage

{ \bf 2. The new model of heat conduction and investigation of its local groups of symmetry.}

Let's change the equation (3) and have it presented as
$$ 
\vec{J}=-\vec{i}\lambda(\frac{\partial T}{\partial x}-
a\cdot arctg(\frac{1}{a}\frac{\partial T}{\partial x})),\eqno(5)   
$$
where $ a $ - parameter, which has dimension of a gradient of temperature. 
From equation (5) it is clear, that flow
$ \vec{J} $ on the absolute value 
never exceeds the flow described in equation (3) more than   
$ 0.5 a\pi $ and relative value of the flows differences 
approaches zero, when temperature gradient increases tending to infinity 
or this gradient being close to zero.
Equation (5) and equation (2) will give the new equation 
of heat conduction:
$$
\frac{\partial T}{\partial t}=D_T
\frac{(\partial T/\partial x)^2}{(\partial T/\partial x)^2+a^2}
\frac{\partial^2T}{\partial x^2}.\eqno(6)
$$

Equation (6) is invariant concerning the local groups of symmetry:
$$ X_1=\partial_t,  X_2=\partial_x,  X_3=\partial_T,  
X_4=2t\partial_t+x\partial_x+T\partial_T.\eqno(7) 
$$
Using invariants 
$ T/\sqrt{t}=const $ and $ x/\sqrt{t}=const $
generated by the operator of dilatation $ X_4 $  
it is possible to obtain replacement of variables
$ z=x/\sqrt{t}, $ 
$ \quad T(x,t)=\sqrt{t}f(z).  $ 
Substituting them into equation (6), we obtain the ordinary 
differential equation for function $ f(z): $
$$
2D_T\frac{\partial^2f}{\partial z^2}(\frac{\partial f}{\partial z})^2=
(f-z\frac{\partial f}{\partial z})
((\frac{\partial f}{\partial z})^2+a^2).\eqno(8)
$$

{ \bf 3. Numeric solution of equation (8).}

Consider boundary value problem for equation (6) with initial condition
$ T(t=0,x)=0, $ boundary conditions 
$ T(t,x=0)=B\sqrt{t}, {\partial T(t, x=0)}/{\partial x}=C, $
$ x\ge 0 $ and assume, that $ D_T=1. $

At reduction of equation (6) to equation (8) boundary conditions 
will be transformed to a kind of 
$ f(0)=B, {\partial f(0)}/{\partial z}=C. $
In figure 1 results of the numerical solution of the equation (8) 
at $ B=1 $, $ \quad C=-1 $, $ \quad a^2=0.001, \quad z=0...5 $ 
are shown. Velocity of movement of heat front being 
$ V_0=V_0(t) $ it is possible to obtain the following:
$$
f(z=z_0)=0 \to z_0=x_0/\sqrt{t} \to V_0=dx_0/dt=z_0/(2\sqrt{t}).
$$
Function $ f = f(z) $ accepts zero value at 
$ z\approx 1.55 $ 
and, consequently, velocity of movement of heat front  will be 
$ V_0\approx 1.55/(2\sqrt{t}). $

This solution qualitatively precisely describes the process of heat conduction 
in real physical media.

{ \bf 4. Example of non-local symmetry of the equation (6).}

Equation (6) by means of non-local transformation
$  H(t,x)={\partial T}/{\partial x}  $
can be transformed to a kind of:
$$ 
\frac{\partial H}{\partial t}=\frac{\partial}{\partial x}
\left(\frac{D_T(\partial H/\partial x)^2}{(\partial H/\partial x)^2+a^2}
\frac{\partial H}{\partial x}\right).
$$
This is a nonlinear equation of heat conduction, in which the coefficient of heat 
conduction is function from gradient of temperature.

{ \bf 5. Quasilinearity of equation (6).}

Numerical researches of equation (8) have shown,that its solutions have physical sense at 
$ 0<a\le 0.1 $. It is possible to assume, that this parameter and in the initial equation
(6) also will accept only such small values. Then equation (6) can be presented as a system 
of the equations:
$$
\frac{\partial T}{\partial t}\approx D_T\frac{\partial^2T}{\partial x^2},\quad
\mbox{ for } \left(\frac{\partial^2T}{\partial x^2}\right)^2>>a^2,\eqno(9)
$$
$$
\frac{\partial T}{\partial t}\approx 0,\quad 
\mbox{ for } \frac{\partial^2T}{\partial x^2}\to 0.\eqno(10)
$$
Equations (9) and (10) are close to a linear kind and, therefore, 
equation (6) can be named as the quasilinear equation of heat conduction.

\vspace*{2mm}

\noindent
{ 
\footnotesize
1.Tikhonov A.N., Samarsky A.A. Equations of mathematical physics. -- Moscow: Nauka, 1966. -- 660p. (in Russian) \\
2.Lykov A.V. Theory of heat conduction. -- Moscow: Vyshaya shkola, 1967. -- 559p. (in Russian) \\
}

\epsfig{file=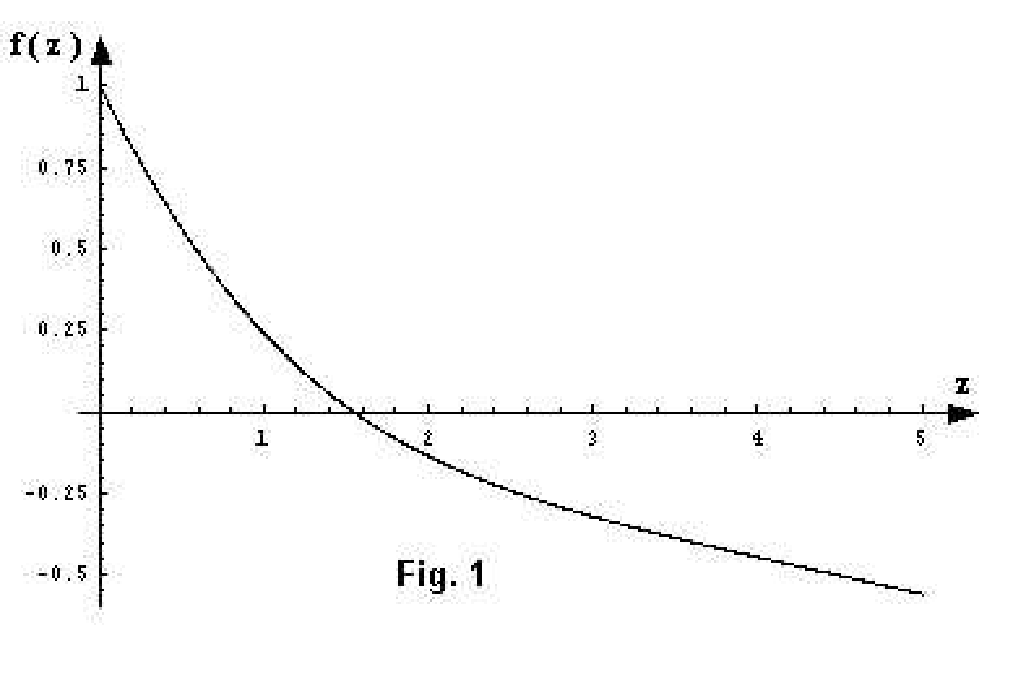, width=14cm, height=10cm.}  

\end{document}